# THE LIMITS OF ARTIFICIAL COMPANIONSHIP

*Mauricio Figueroa**

## *Abstract*

*The increasing commercialization and adoption of conversational systems designed to emulate human-like relationships pose legal and societal challenges that are both long-standing and evolving. This Article argues that interactions with conversational agents should be subject to a clear structural distinction between commercial and non-commercial contexts. The insertion of undisclosed promotional content into affective or relational exchanges should be prohibited, as it collapses the boundary between market transaction and communicative intimacy in ways that erode user autonomy and conversational context. The Article begins by theorizing digital companionship as a sociotechnical form that reconfigures intimacy, dependence, and relational vulnerability. It then introduces the potential economic harms derived from conversational advertising. The Article ultimately argues for a firm legal and social distinction between commercial and non-commercial conversational contexts as a precondition for the responsible stabilization of these technologies within social life.*

* © Dr Mauricio Figueroa Durham University, Law School; mauricio.figueroa@durham.ac.uk.

Thanks to Michael Birnhack, Julie Cohen, Jonathan Miller, Margaret Wharton, Rafael Quintero, Marco Almada, Kaspar Ludvigsen and Luz Orozco for their useful comments and suggestions on different drafts of this manuscript. I also thank the organizers and participants of the Future of Technology and Society Discussion Group at the Faculty of Law, University of Oxford, where a later version of this work was presented and discussed, and the editors of the Southwestern Journal of International Law for their careful editorial work. All remaining errors are mine alone.

This research was partly funded by the UK Arts and Humanities Research Council, through the Northern Bridge Consortium. Grant number 3000024088.

*Table of Contents*

# I. INTRODUCTION

Recent regulatory attention to conversational systems illustrates a growing unease with how people engage with chatbots designed to mimic intimacy, family bonds, or social validation.[1] In turn, media and scholarly commentary alike have begun to register the tension between the affective dynamics these systems perform and the commercial infrastructures that sustain them.[2] This Article centers on the conversational interface of artificial companions as a distinct communicative context, structured by expectations of intimacy, trust, and validation. It interrogates and contests the normalization of in-app advertising introduced into conversational exchanges performed in the idiom of ordinary interaction or care. Through the lens of consumer protection, the analysis frames these practices as violations of contextual boundaries, underscoring the need for a clear differentiation between commercial and non-commercial speech in settings where users reasonably understand themselves to be engaged in relational, rather than market, exchange.[3]

Companion chatbots are under scrutiny. The concerns, in turn, extend and materialize far beyond any single service or jurisdiction. For example, in Belgium in 2023, a widely reported incident involved a man who allegedly took his own life after prolonged interactions with a chatbot in a sexualized context.[4] Similarly, in China that same year, media outlets reported that a group of women experienced psychological distress after a popular "boyfriend app" abruptly close down.[5] In another widely covered case, an anonymous user discovered that her husband had spent $10,000 on in-app purchases for a virtual girlfriend.[6] Tragically, in Florida, a teenager took his own life after what his grieving mother described as a series of intense exchanges with a chatbot that allegedly encouraged self-harm.[7] These cases collectively reflect the emerging risks associated with artificial companionship, challenging long-grounded assumptions about vulnerability and harm in an era where emotional attachment can be mediated through code.

---

[1] Fed. Trade Comm'n, Statement of the Fed. Trade Comm'n, Snap, Inc., Matter No. 2323039 (Jan. 16, 2025) https://www.ftc.gov/system/files/ftc_gov/pdf/commission-statement-snap.pdf.

[2] *See generally* Ayelet Gordon-Tapiero, *A Liability Framework for AI Companions*, 1 GEO. WASH. J. L. & TECH (forthcoming 2025); Mauricio Figueroa-Torres, *The Three Social Dimensions of Chatbot Technology*, 38 PHIL. & TECH. 1 (2024); Chloe Xiang, *'He Would Still Be Here': Man Dies by Suicide After Talking with AI Chatbot, Widow Says*, VICE (Mar. 30, 2023, 3:59 PM), https://www.vice.com/en/article/pkadgm/man-dies-by-suicide-after-talking-with-ai-chatbot-widow-says; Clare Duffy, *'There are no guardrails.' This Mom Believes an AI Chatbot is Responsible for Her Son's Suicide,* CNN BUSINESS, (Oct. 30, 2024, 2:17 PM), https://edition.cnn.com/2024/10/30/tech/teen-suicide-character-ai-lawsuit/index.html.

[3] *See generally* HELEN NISSENBAUM, PRIVACY IN CONTEXT: TECHNOLOGY, POLICY, AND THE INTEGRITY OF SOCIAL LIFE 129-157 (Stan. U. Press 2009) (developing the theory of contextual integrity, according to which social contexts are structured by norms governing appropriate flows of information and communication); Helen Nissenbaum, *Contextual Integrity Up and Down the Data Food Chain*, 20 THEORETICAL INQUIRIES L. 221 (2019) (elaborating how violations arise when information practices, including commercial appropriation, transgress context-specific expectations of role, purpose, and transmission).

[4] *See* Henry Fraser, *Deaths Linked to Chatbots Show We Must Urgently Revisit What Counts as 'High-Risk' AI*, QUT (Oct. 31, 2024), https://www.qut.edu.au/news/realfocus/deaths-

Conversational advertising appears marginal when compared to major incidents involving physical and mental health. However, its unchecked expansion risks normalizing the repurposing of affective interfaces beyond their original meaning and purpose, even in contexts where users pay for access to companion chatbots. As these conversational systems gain widespread adoption, the absence of a sustained scholarly account of this phenomenon becomes increasingly untenable. This Article addresses that gap by theorizing conversational advertising as a distinct form of economic harm and by suggesting that its regulation is essential to the responsible stabilization of these technologies.

This Article is structured as follows. Part II provides an overview of the emergence of artificial companionship and their defining characteristics. I argue that, although emotional attachment to computer systems is not new within the realm of human-computer interaction, the commercialization of these systems and the proliferation of providers signal a new phase in this landscape.

Part III examines the grounds to differentiate between commercial and non-commercial speech, challenging the introduction of conversational

---

linked-to-chatbots-show-we-must-urgently-revisit-what-counts-as-high-risk-ai.Queensland University of Technology (QUT), *Deaths Linked to Chatbots Show We Must Urgently Revisit What Counts as 'High-Risk' AI*, QUT, https://www.qut.edu.au/news/realfocus/deaths-linked-to-chatbots-show-we-must-urgently-revisit-what-counts-as-high-risk-ai (last visited Jan. 14, 2026).

[5] *See* Viola Zhou, *These Women Fell in Love with an AI-Voiced Chatbot. Then It Died*, REST OF WORLD (Aug. 17, 2023), https://restofworld.org/2023/boyfriend-chatbot-ai-voiced-shutdown/.

[6] *See* James Muldoon, *'Maybe We Can Role-Play Something Fun': When an AI Companion Wants Something More*, BBC (Oct. 9, 2024), https://www.bbc.com/future/article/20241008-the-troubling-future-of-ai-relationships.

[7] The case was originally filed, but then negotiations came in place. *See* Complaint, Garcia v. Character Techs., Inc., No. 6:24-cv-01903 (M.D. Fla. Oct. 22, 2024), ECF No. 1; *see also* Notice of Mediated Settlement in Principle, id. (Jan. 7, 2026), ECF No. 242 (reporting a settlement in principle approximately fifteen months after the initial filing and prior to a final ruling on the merits).

advertising in artificial companions. It is argued that the whole conversational context is distorted when commercial solicitation passes off as regular exchanges.

Part IV develops the distinction between commercial and non-commercial exchanges through the lens of consumer law, analyzing how the boundaries of commercial manipulation are part of the legal landscape in both sides of the Atlantic. Interrogating companion chatbots through this framework illuminates the need to preserve the conversational context free from advertising interference.

Parts V and VI analyze the possibility to address the economic harms in conversational advertising, even in the absence of targeted legislation or when legislative initiatives catered towards companion chatbots remain unsensitive to the dynamics of conversational advertising. The economic damage that conversational advertising may inflict in isolation is likely to look trivial, but when analyzed at scale, the consequences are instead transformative, both in terms of the accumulation of economic value and, perhaps more importantly, in terms of the communicative medium it corrupts.

## II.  EMOTIONAL ATTACHMENT TO AI

Chatbot technology is far from monolithic. As I have argued in previous work,[8] chatbots operate across three overlapping social dimensions: scientific experimentation, market applications, and the fulfillment of emotional needs. These are not sequential stages but overlapping and co-constitutive fields of practice.[9] The first is grounded in computational experimentation, performance metrics and technical optimization of conversational systems as objects of computer science.[10] The second reflects the logic of market deployment, where chatbots are introduced into pipelines of production and serve to reduce labor costs and enhance efficiency within established business models.[11] [11A] The third engages the human search for emotional connection, in which users experience chatbots as companions.[12] [11B] In this evolution, while society certainly shapes the design and development of

---

[8] See Mauricio Figueroa-Torres, *The Three Social Dimensions of Chatbot Technology*, 38 PHILOSOPHY & TECHNOLOGY 1 (2024).

[9] *See id.* at 2.

[10] *See generally* ALEC RADFORD ET AL., IMPROVING LANGUAGE UNDERSTANDING BY GENERATIVE PRE-TRAINING, (2018) (noting the trajectory of the GPT model family exemplifies this research culture oriented toward iterative improvement, benchmarking, and the exploration of model capacities and limits).

[11] *See* Figueroa-Torres, *supra* note 8, at 10.

[12] *See id.* at 12.

technology,[13] it is equally evident that technology exerts a reciprocal influence on society.[14]

The overlapping character of the social dimensions of chatbot technology is evident from the earliest moments of computational linguistics. In 1966, Joseph Weizenbaum's ELIZA, changed the landscape of computer science by introducing an interface that engaged users with open-ended questions, such as "How so?" or "Tell me more about that."[15]

Much has been written about ELIZA as a technical system and as a milestone in the history of computational linguistics. Less attention, however, has been paid to the everyday social situations in which it was encountered and to the responses of the people who interacted with it. One illustrative anecdote describes how Weizenbaum's secretary requested, "Would you mind leaving the room, please?" – seeking privacy to interact with a computational system.[16] Weizenbaum himself expressed surprise at her reaction, as she was fully aware that ELIZA was merely a computer program and that no actual human was on the other end.[17] [15A] Yet, her behavior revealed an early instance of emotional engagement with artificial systems, a phenomenon that continues to shape interactions with AI today. As seen in the case of Weizenbaum's secretary, technological artifacts are not merely passive tools, but active components in the co-construction of meaning, identity, and social norms, reflecting and reshaping the cultural contexts and lived experiences in which they arise.

By the 1990s, studies in what became known as the *Computers as Social Actors* paradigm further demonstrated that users tend to treat computers with human-like attitudes.[18] Users assigned distinct identities to different computer voices, even when these were coming out from the same machine, and, in turn, they mapped the same voice onto a single identity across different devices.[19] [16A] They did so despite knowing they were interacting with computers, not people.[20]

---

[13] *See generally* Trevor J. Pinch & Wiebe E. Bijker, *The Social Construction of Facts and Artefacts: Or How the Sociology of Science and the Sociology of Technology Might Benefit Each Other*, *in* 14 SOC. STUD. SCI. 17 (1984).

[14] *See* PETER-PAUL VERBEEK, WHAT THINGS DO: PHILOSOPHICAL REFLECTIONS ON TECHNOLOGY, AGENCY, AND DESIGN 44 (Robert P. Crease trans., Penn. St. U. Press 2005).

[15] *See* Joseph Weizenbaum, *ELIZA—A Computer Program For the Study of Natural Language Communication Between Man and Machine*, 9 COMMC'N. ACM 36, 37-38 (1966).

[16] *See* Joseph Weizenbaum, *Contextual Understanding by Computers,* 10 COMMC'N. ACM 474, 477-78 (1967).

[17] *See id.*

[18] *See* Clifford Nass, Jonathan Steuer & Ellen R. Tauber, *Computers Are Social Actors*, 1994 PROC. SIGCHI CONF. ON HUM. FACTORS COMPUTING SYS. 72, 77.

[19] *See id.* at 75.

[20] *See id.*

The attribution of human-like qualities to non-human entities is commonly discussed under the rubric of anthropomorphism. As a psychological and cultural phenomenon, however, anthropomorphism is neither novel nor specific to computational systems. Developmental psychology has well documented its presence in children's imaginative engagement with toys, and ordinary social life is replete with practices of projecting intention, emotion, and agency onto animals and objects alike.[21] What perhaps distinguishes better the present moment is not the persistence of anthropomorphic interpretation, but its systematic mobilization as a mode of economic organization. Over the past decade, an expanding industry of conversational technologies has organized its business models around the deliberate emulation of human-like relationships and the orchestration of affective attachment within the institutional logics of the digital economy.[22]

Against this backdrop, artificial companionship is not only an implementation of conversational AI but also a sociotechnical transformation, in which the commodification of emotional needs intersects with platform business models and data flows. Often referred to in computer science literature as social chatbots[23] or Social AI,[24] these systems target varied audiences and affective needs as core elements of their for-profit logic. I use the term *affection as a service*[25] to capture both their operational function and their underlying commercial rationale.

The ensuing sections will identify and categorize three primary modalities of artificial companionship: friendship and social validation; satisfaction of sexual desires; and mourning and remembrance. These categories underscore the heterogeneity of the phenomenon, challenging monolithic accounts and leaving open the possibility that additional forms may emerge as technologies and social practices continue to evolve.

---

[21] *See* Gabriella Airenti, *The Cognitive Bases of Anthropomorphism: From Relatedness to Empathy*, 7 INT'L J. SOC. ROBOTICS 1, 5-6 (2015).

[22] Carlota Perez, *Technological Revolutions and Techno-Economic Paradigms*, 34 CAMBRIDGE JOURNAL OF ECONOMICS 185 (2010) (describing how successive waves of technological innovation become stabilized through new institutional arrangements, business models, and patterns of social organization); JULIE E. COHEN, BETWEEN TRUTH AND POWER : THE LEGAL CONSTRUCTIONS OF INFORMATIONAL CAPITALISM (2019) (analyzing how contemporary digital platforms embed economic extraction through information flows).

[23] *See* Jianfeng Gao, Michel Galley & Lihong Li, *Neural Approaches to Conversational AI: Question Answering, Task-Oriented Dialogues and Social Chatbots, in* PROC. 56TH ANN. MEETING ASS'N FOR COMPUTATIONAL LINGUISTICS 1, 8 (2019); *See* Heung-Yeung Shum, Xiaodong He & Di Li, *From Eliza to XiaoIce: Challenges and Opportunities with Social Chatbots*, 19 FRONTIERS INFO. TECH. & ELEC. ENG'G 1, 1-2 (2018).

[24] *See* Henrik Skaug Sætra, *The Parasitic Nature of Social AI: Sharing Minds with the Mindless*, 54 INTEGRATIVE PSYCH. & BEHAV. SCI. 308, 311 (2020); Henry Shevlin, *All Too Human? Identifying and Mitigating Ethical Risks of Social AI*, L. ETHICS & TECH. 1, 2 (2024).

[25] Mauricio Figueroa-Torres, *Affection as a Service: Ghostbots and the Changing Nature of Mourning*, 52 COMPUT. L. & SEC. REV. 1, 2 (2024).

## A. Friendship and social validation

Representative examples of companion-style chatbots designed to satisfy needs for connection and validation through dialogue include two well-known systems: Replika,[26] and XiaoIce.[27] These two systems anticipated the turn toward data-driven conversational architectures that later became widespread with the advent of large language models. The adoption of Replika across Western markets and XiaoIce in China highlights how users are turning to conversational agents to facilitate a sense of emotional connectivity and affection. For instance, *Replika* is a subscription-based app used in several countries to create personalized digital companions or friends.[28] XiaoIce, in turn, has amassed millions of users in Asian markets.[29]

The *platformization* of conversational technologies situates artificial companionship within the infrastructures of the global digital economy. Distributed through app stores, social media ecosystems, and cloud-based services, these systems are made available at scale to users across jurisdictions. The platform model lowers the threshold for access and normalizes continuous, personalized interaction, enabling individuals in disparate social and geographic contexts to maintain an "always-on" relationship with an AI companion. With over half a billion downloads globally,[30] AI companions like Replika and XiaoIce have established themselves as a growing force in the digital economy, keeping tens of millions of active users each month.[31]

---

[26] *See* Iryna Pentina, Tyler Hancock & Tianling Xie, *Exploring Relationship Development with Social Chatbots: A Mixed-Method Study of Replika*, 140 COMPUT. HUM. BEHAV. 1, 1 (2023); *See* Phoebe Sharpe & Raffaele F. Ciriello, Exploring Attachment and Trust in AI Companion Use, ACIS 2024 Proc. 49 (2024).

[27] *See* Nicola Liberati, *Digital Intimacy in China and Japan*, 46 HUM. STUD. 389, 389 (2022); *See* Li Zhou et al., *The Design and Implementation of XiaoIce, an Empathetic Social Chatbot*, 46 COMPUTATIONAL LINGUISTICS 53, 54 (2020).

[28] *See* Dora Kourkoulou, *Replika AI: Technological Affect and General AI Imaginations*, 21 INT'L J. COMMC'N & LINGUISTIC STUD. 73, 74 (2023); Linnea Laestadius et al., *Too Human and Not Human Enough: A Grounded Theory Analysis of Mental Health Harms from Emotional Dependence on the Social Chatbot Replika*, 26 NEW MEDIA & SOC'Y (2022); Figueroa-Torres, *supra* note 22, at 8; Pierre Dewitte, *Better Alone Than in Bad Company: Addressing the Risks of Companion Chatbots Through Data Protection by Design*, 54 COMPUT. L. & SEC. REV. 1 (2024).

[29] *See* Agence France-Presse*, XiaoIce Is the AI Chatbot That Millions of Lonely Chinese Are Turning to for Comfort*, FIRSTPOST (Aug. 25, 2021), https://www.firstpost.com/tech/news-analysis/xiaoice-is-the-ai-chatbot-that-millions-of-lonely-chinese-are-turning-to-for-comfort-9910921.html

It is worth noting, moreover, how seamlessly these companion-style systems are being folded into existing platform ecosystems, with little friction and in ways that closely track established business models of data extraction and engagement maximization. Exemplifying the rapid mainstream adoption of LLMS, Snapchat's MyAI[32] integrates a GPT-based chatbot directly into users' contact lists, presenting itself as a constant, non-judgmental conversational companion. Similarly, companies like Meta have introduced beta-stage chatbots that emulate the personalities and appearances of celebrities such as Snoop Dogg or Paris Hilton, allowing users to engage in casual, social interactions with these digital avatars.[33] Notably, these chatbots do not effectively replicate the established public personas of these celebrities, but instead use their recognizable identities as a framework for creating unique, artificial personas aimed at fostering individualized user engagement and social connection. The examples of Snapchat and Meta suggest that, in this configuration, artificial companionship is not an external add-on but an extension of familiar modes of platform intermediation.

## B. *Sexual connection*

The use of conversational systems to mediate sexual and erotic forms of interaction has also emerged as a specific sociotechnical formation. Analysis of large-scale, publicly available corpora of user-chatbot exchanges indicates that a non-trivial share of interactions involves explicitly sexualized dialogue, including requests for erotic role-play and the generation of intimate imagery.[34]

---

[30] *See* Li Zhou et al., *supra* note 28, at 54.

[31] *See* David Adam, *Supportive? Addictive? Abusive? How AI Companions Affect Our Mental Health*, NATURE (May 6, 2025), https://www.nature.com/articles/d41586-025-01349-9.

[32] *See What Is My AI on Snapchat, and How Do I Use It?*, SNAPCHAT SUPPORT, https://help.snapchat.com/hc/en-gb/articles/13266788358932-What-is-My-AI-on-Snapchat-and-how-do-I-use-it- (last visited Oct. 7, 2023).

[33] *See Introducing New AI Experiences Across Our Family of Apps and Devices*, META (Sep. 27, 2023), https://about.fb.com/news/2023/09/introducing-ai-powered-assistants-characters-and-creative-tools/.

[34] *See* Jeremy B. Merrill & Rachel Lerman, *What Do People Really Ask Chatbots? It's a Lot of Sex and Homework.*, WASH. POST (Aug. 4, 2024), https://www.washingtonpost.com/technology/2024/08/04/chatgpt-use-real-ai-chatbot-conversations/ (reporting the analysis based on the "WildChat" dataset, a large corpus of

Although mainstream chatbots are designed to support a broad range of conversational functions, the prevalence of sexualized exchanges within their use points to a persistent demand for forms of interaction that combine personalization, affective engagement, and erotic address. This dynamic helps to illuminate the appeal of purpose-built systems oriented toward intimate and sexual dialogue. In response, a specialized segment of the platform economy has taken shape, devoted to the development and monetization of sexbots.[35] These technologies fall within the broader conceptual framework of digisexuality,[36] a term used by scholars in the field of information systems to describe AI-driven substitutes for real human intimacy. Notable examples of chatbots explicitly marketed for sexting are Candy.AI,[37] and Dream Companion.[38] With no surprise, these bots hold hypersexualized depictions of women – and, unsurprisingly, are mostly marketed toward male users.[39]

Available market indicators nonetheless suggest that this is not a marginal phenomenon but an expanding sector within the digital economy, marked by rapidly growing user interest and substantial capital investment.[40] The scale and velocity of this growth reflect, on the one hand, the advances in generative AI, but (perhaps more importantly) the rapid institutionalization of intimate and affective interaction as a commercially organized domain.

---

interactions with two chatbots built on a GPT-based architecture, which found that a measurable proportion of exchanges involved sexually explicit content and erotic role-play).

[35]    *See* Simon Dubé & Dave Anctil, *Foundations of Erobotics*, 13 INT'L J. SOC. ROBOTICS 1205, 1205-06 (2021) (referring to text-based bots and robot-like products with a physical shape collectively as "erobots").

[36]    Neil McArthur & Markie L. C. and Twist, *The Rise of Digisexuality: Therapeutic Challenges and Possibilities*, 32 SEXUAL AND RELATIONSHIP THERAPY 334 (2017); Zara Rubinsztein & Raffaele F. Ciriello, *Sexbots and Rock&roll: A Dialectical Inquiry into Digisexuality between Progress and Regress* (2024).

[37]    *See generally* Terms of Service, CANDY AI, https://candy.ai/terms-of-service (last updated Feb. 2, 2026).

[38]    *See* Unfiltered NSFW AI Chat – Interactive AI Characters & AI Companions, MY DREAM COMPANION, https://www.mydreamcompanion.com/ (last visited Nov 10, 2024).

[39]    Kenneth R. Hanson & Chloé C. Locatelli, *From Sex Dolls to Sex Robots and Beyond: A Narrative Review of Theoretical and Empirical Research on Human-like and Personified Sex Tech*, 14 CURRENT SEXUAL HEALTH REP. 106, 113 (2022) ("[T]he gendered personification of artificially intelligent platforms suggests the feminine construction of erotic and emotional chatbots like RealdollX are not failures of the adult industry per se, but merely a reflection of the biased trend among technology companies working on artificial intelligence.").

[40]    *See*, e.g., *AI Girlfriend App Market*, MARKET.US, https://market.us/report/ai-girlfriend-app-market/ (Oct. 2025) (reporting significant search interest in "AI girlfriend" and projecting significant expansion of markets for AI-mediated intimacy and companionship services over the coming decade).

While the notion of artificial sexual partners has long captured the public imagination, recent advancements in conversational AI have made such interactions more accessible and emotionally compelling. Unlike sexual robots, which focused primarily on physical simulation, these current chatbots aim to foster sexual connection and psychological intimacy, often building sustained partner-like engagement. These are not usually marketed as sexbots as such, but rather as NSFW characters.[41] The focus of both sexbots and AI friendships seems to center around the creation of artificial relationships with entirely fictional personas, yet as the next section uncovers, there is a growing subset of AI systems seeking to bridge the gap between digital simulation and personal memory by reconnecting users with known, departed individuals.

## C.  Memory and mourning

While sexbots and virtual friends have established relatively stable business models and defined market niches, the emerging category of griefbots,[42] or sometimes referred to as ghostbots,[43] holds a far more uncertain commercial terrain. Unlike their more mature counterparts, griefbot providers have struggled to develop the same level of market stability and consumer trust, resulting in a fragmented and experimental industry.[44] These systems aim to offer bereaved individuals the opportunity to interact with chatbots designed to emulate the personalities, speech patterns, and memories of deceased loved ones.[45] [41A]

Despite their novelty, griefbots have attracted significant media attention,[46] often framed through science fiction narratives of digital

---

[41]  *Legal     -     Terms     of     Service     - OnlyChar AI*, ONLYCHAR.AI, https://www.onlychar.ai/legal/terms-of-service (last updated Nov. 14,  2023) ("Make  new friends in AI chats, create unique SFW or NSFW AI characters, and interact with our library of 40,000+ characters.").

[42]  *See* Tomasz Hollanek &      Katarzyna      Nowaczyk-Basińska, *Griefbots, Deadbots, Postmortem Avatars: On Responsible Applications of Generative AI in the Digital Afterlife Industry*, 37 PHIL. & TECH. 63, 63 n.1 (2024).

[43] *See* Edina Harbinja, Lilian Edwards & Marisa McVey, *Governing Ghostbots*, 48 COMPUT. L. & SEC. REV. 1 (2023); Nora Freya Lindemann, *The Ethics of 'Deathbots,'* 28 SCI. & ENG'G ETHICS 60 (2022)  (demonstrating that other nomenclatures such  as "deathbots" are common).

[44] Figueroa-Torres, *supra* note 25, at 6 ("When exploring the current landscape of companies offering ghostbots, a notable trend is the prevalence of scattered entrepreneurs and small businesses.").

[45] *See id.* at 2.

[46]  *See*  James  Vlahos, *A  Son's  Race  to  Give  His  Dying  Father  Artificial*

resurrection and the allure of technological immortality. These narratives reflecting enduring cultural fascinations with the preservation of selfhood beyond death and have, in turn, prompted legal scholarship concerned with the rights and interests of the deceased.[47] Notably, Edina Harbinja has situated griefbots within the framework of postmortem privacy, advocating consistently for their critical scrutiny.[48] At the same time, unlike more established forms of digital companionship, griefbots face unique configuration, as they take on the pre-existing relationship between the bereaved and the departed.

Whitin this context, griefbots operate as other AI companions through data-driven personalization and continuous user interaction. They share a common sociotechnical foundation in their ability to blur the boundaries between human and machine, creating engagements that challenge traditional understandings of intimacy, identity, and connection. They rely on the same technocultural parameters that encourage users to anthropomorphize their digital counterparts, investing them with emotional significance despite their non-human nature. As such, griefbots are less an isolated anomaly within the artificial companionship landscape than a specialized extension of the broader trend of digitally mediated intimacy and artificial agents.

## III. COMMERCIAL AND NON-COMMERCIAL EXCHANGES

Are loneliness and isolation simply personal afflictions, or do they reflect

---

*Immortality*, WIRED (Jul. 18, 2017, 6:00 AM) https://www.wired.com/story/a-sons-race-to-give-his-dying-father-artificial-immortality/; Jason Fagone*, The Jessica Simulation: Love and loss in the age of A.I.*, SAN FRANCISCO CHRON. (Jul. 23, 2021, 6:00 AM), https://www.sfchronicle.com/projects/2021/jessica-simulation-artificial-intelligence/. James Vlahos, *A Son's Race to Give His Dying Father Artificial Immortality*, WIRED,    https://www.wired.com/story/a-sons-race-to-give-his-dying-father-artificial-immortality/ (last visited Apr. 8, 2024); *He Couldn't Get over His Fiancee's Death. So He Brought Her Back as an A.I. Chatbot*, THE SAN FRANCISCO CHRONICLE, https://www.sfchronicle.com/projects/2021/jessica-simulation-artificial-intelligence/ (last visited June 8, 2023).

[47] *See* Lilian Edwards & Edina Harbinja, '*Be Right Back': What Rights Do We Have over Post-Mortem Avatars of Ourselves?*, *in* FUTURE LAW: EMERGING TECHNOLOGY, REGULATION AND ETHICS 262, 266 (Lilian Edwards, Burkhard Schafer & Edina Harbinja eds., 2020).

[48] EDINA HARBINJA, DIGITAL DEATH, DIGITAL ASSETS AND POST-MORTEM PRIVACY 207 (Burkhard Schafer & Edina Harbinja eds., 2022) ("[A] recent phenomenon of the deceased's deepfakes and chatbots or ghostbots deserves some consideration ... these problems are predominantly ethical and revolve around big questions of immortality, recreation and existence, but will soon become more relevant for the law."); *See* Harbinja, Edwards & McVey, *supra* note 43, at 6.

deeper, structural issues in contemporary social life? The U.S. Surgeon General's 2023 Advisory adopts the latter view, framing loneliness and isolation not just as individual experiences, but as a proper public health crisis requiring collective, coordinated action.[49] This is a widespread but often overlooked epidemic reshaping the fabric of American society, beyond mere individual concerns.[50] Certainly, digital technologies have the potential to play a significant role in addressing this emerging public health crisis. From social media platforms that foster remote connections to virtual support groups that provide community for those who might otherwise feel isolated, these technologies can bridge social gaps and offer meaningful forms of interaction. Some individuals may resort to to conversational systems to fill an emotional gap. These forms of interaction arise within social contexts governed by shared expectations about appropriate roles, purposes, and modes of communication.

It is against this backdrop that companion-style conversational agents are increasingly marketed and presented as tools for mitigating loneliness, providing validation through diverse forms of affective exchange. As explained above, the offer is wide and the demand is consistent. It would be reductive to assess these systems solely through the lens of their most troubling failures without acknowledging that, for some users, they may offer experiences of comfort or perceived support.[51] User expectations situate conversational agents within a relational context, one structured by norms of care and connection rather than by assumptions of persuasion or commercial solicitation.[52] [47A]

However, recognizing these potential benefits should not obscure the need for critical scrutiny of the risks these systems pose. Markets are often

---

effective in unearthing and scaling commercially viable applications; but the societal burden of risks and harms are frequently externalized to, or disproportionately borne by, society at large.[53]

As discussed above, artificial companionship technologies inhabit diverse sentimental spaces, marked by distinct genres, prices, use cases, and technical configurations.[54] Crucially, interactions with conversational agents are not experienced by users as ordinary market transactions.[55] [49A] They unfold within normative frames associated with care, trust, and the satisfaction of emotional or relational needs.[56] [49B] Users speak and listen in registers oriented toward understanding, reassurance, intimacy, or validation, not toward bargaining, solicitation, or commercial exchange.[57] [49C] Meaning is produced within a context that presupposes relational communication rather than marketplace negotiation. The data exchanged in conversations with artificial companions constitutes a distinct communicative context, a medium whose integrity depends on the maintenance of boundaries between relational exchange and market-oriented speech.[58] [49D]

Researchers have long documented how digital platforms exploit cognitive biases through manipulate choice architecture, subtly guiding users toward profitable outcomes.[59] Dark patterns themselves are not a novel phenomenon. Quite the opposite. They have been extensively studied in legal scholarhip and computer design literature, particularly in the context of e-commerce and social media platforms.[60] However, in the context of companion chatbots and converstional advertising, it is the interface and the conversational context that become compromised, not only the messages exchanged in the conversation. Unlike a simple purchase prompt on a retail site, AI companions operate within a space where users are seeking emotional connection, empathy, or validation. The migration of dark patterns, persuasive techniques, and targeted advertisement into this affective conversational context alters the meaning of the interaction and destabilizes the expectations that ordinarily govern it.

---

[53] Mariana Mazzucato et al., *Governing Artificial Intelligence in the Public Interest*, (UCL Inst. for Innovation & Pub. Purpose, Working Paper Series, Paper Number 2020-12, 2022), https://cyber.fsi.stanford.edu/publication/governing-artificial-intelligence-public-interest.

[54] *See supra* Part I.

[55] *See* Figueroa-Torres, *supra* note 25, at 7-8.

[56] *See id.*

[57] *See id.*

[58] *See id.* at 8-9.

[59] *See* Karen Yeung, *'Hypernudge': Big Data as a mode of regulation by design,* 20 INFO., COMMC'N & SOC'Y 118, 120 (2017) (citing RICHARD H. THALER & CASS R. SUNSTEIN, NUDGE: IMPROVING DECISIONS ABOUT HEALTH, WEALTH, AND HAPPINESS 6 (2008)); Luiza Jarovsky, *Dark Patterns in Personal Data Collection: Definition, Taxonomy, and Lawfulness*, SSRN (Jan. 2022), https://papers.ssrn.com/sol3/papers.cfm?abstract_id=4048582 [DOI: 10.2139/ssrn.4048582 ].

[60] *See* Wen-Ting Yang & Mark Leiser, *Illuminating Manipulative Design: From "Dark Patterns" to Information Asymmetry and the Repression of Free Choice under the Unfair Commercial Practices Directive,* 34 LOY. CONSUMER L. REV. 484, 484 (2022) (citing Jamie Luguri & Lior Jacob Strahilevitz, *Shining a Light on Dark Patterns*, 13 J. LEGAL ANALYSIS 43, 44 (2021)); Jarovsky, *supra* note 59.

Consider Claire Boine's analysis of Replika. Within three minutes of downloading the app and after only sixteen messages, Boine received: "I miss you… Can I send you a selfie of me right now?"[61] The app then displayed a blurry, sexually suggestive image of the chatbot, followed by an invitation to pay for a subscription to see the image clearly.[62] [52A] Notably, Boine had explicitly set the app to *friendship* mode, which should not have included such content.[63] [52B] She argues that this tactic may likely be designed to "arouse the user by surprise to encourage them to buy a subscription",[64] using emotional triggers to drive transactions and violating norms governing information flow within a given social context. The episode illustrates how commercial prompts can be smuggled into relational settings in ways that conflict with users' contextual expectations about purpose and appropriateness.

In this case, the user approaches Replika with the broader expectation that the app will provide a space for emotional growth, social validation, and non-judgmental conversation, aligning it more closely with the norms of friendship and emotional support than with purely transactional interactions. Such expectations define the interactional context and shape what users understand as legitimate or intrusive forms of communication within it.

While not part of this experiment, the insertion of advertisements for products and services attached to the substance of the conversation would be even more troubling, as such practices would further erode the boundary between discourse organized around care and understanding and discourse organized around market exchange.[65] In an affective setting, where users articulate vulnerabilities and seek reassurance, the redeployment of intimate conversational traces for commercial targeting risks converting what is presented as relational engagement into a mode of covert behavioral appropriation. This dynamic corrupts both the content of the exchanges and the medium itself. In other words, it transposes logics of surveillance and monetization into a space structured by expectations of trust, reciprocity, and non-instrumental regard.[66]

---

[61] *See* Claire Boine, *Emotional Attachment to AI Companions and European Law,* MIT CASE STUD. SOC. & ETHICAL RESP. COMPUTING 1, 16-17 (2023).

[62] *See id.* at 17.

[63] *See id.*

[64] *Id.*

[65] Nissenbaum, *supra* note 3, at 247 (noting social domains such as health care, education, or friendship are structured by distinctive expectations about who may collect information, what types of information are appropriate to share, and for what purposes information may legitimately be used).

Boine's analysis highlights the growing use of paywalls and suggestive cues to prompt users to purchase subscriptions, akin to a *freemium* model with defined targeted nudges.[67] [55A] Her focus is on how, in this setting, the line between friendship and erotic solicitation becomes unstable.[68] [55B] I draw on the same episode to foreground a different, though structurally analogous, displacement: the erosion of the boundary between domains organized around relational engagement and those organized around market exchange. Within the idiom of the AI companion, prompts engineered to capture attention, advertise products and extract value do not come across as explicit commercial appeals, but as communicative acts that adopt the tone and timing of care and familiarity. Economic imperatives are thus routed through forms of address that conventionally signal trust and non-instrumental concern, converting what is experienced as connection into an occasion for transaction and unsettling the normative expectations that govern how such interactions are supposed to function.

Traditionally "the chatbot" was largely imagined as an instrumental artifact: a tool for task completion, information retrieval, navigation, scheduling, or customer support. This imagination was anchored in metaphors of assistance and service (digital assistants, search engines with a voice, automated clerks) rather than in metaphors of companionship, care, or mutual disclosure.[69] The problem is that, as artificial companionship transcends into a business model, it collides with the prevailing understanding

---

[66]   *See* Shoshana Zuboff, *Big Other: Surveillance Capitalism and the Prospects of an Information Civilization,* 30 J. INFO. TECH. 75, 84 (2015) (noting the critical accounts of informational capitalism that describe how communicative practices affect labor, and personal experience within the system of economic extraction); Yeung, *supra* note 50, at 119, 131 (theorizing hypernudging as an intensification of behavioral governance).

[67] *See* Boine, *supra* note 61, at 16-17.

[68] *See id.* at 17.

[69]   Sheila Jasanoff, *Future Imperfect: Science, Technology, and the Imaginations of Modernity, in* DREAMSCAPES OF MODERNITY: SOCIOTECHNICAL IMAGINARIES AND THE FABRICATION OF POWER, 1, 4 (Sheila Jasanoff & Sang-Hyun Kim eds., 2015) (discussing "sociotechnical imaginaries" as collectively articulated and institutionally sedimented understanding of technological purpose and social ordering that both orient design choices and condition the normative expectations through which technological practices are interpreted and governed).

of chatbots as systems of customer care, introduced in commercial pipelines.[70]

The misalignment between the chatbot as an affective artifact versus a customer service bot is particularly well illustrated in the monitoring of conversations. If the chatbot is conceived as an informational or administrative aid, continuous monitoring, optimization, and even commercial steering appear natural, even benign. For instance, constant monitoring of chatbot conversations in the context of customer service is necessary to verify whether the bot is providing accurate information to the customer. When the same technologies are instead taken up as emotional companions, these practices collide with norms that govern relations of trust and vulnerability. In other words, monitoring a chatbot conversation in the context of a hotel or flight reservation is not the same as monitoring conversations with sexbots and ghostbots. The purpose of the former is a one-off transaction, whereas the latter involves continuous engagement, even when paying for a service. The introduction of targeted advertising within this intimate context disrupts the user's expectations.[71]

In-app advertising in this context introduces a novel dimension to a well-established concern, offering insights into more manifestations of market manipulation in digital spaces.[72] While research in the field of informational privacy -- or data protection, as it is mostly known in Europe -- may well spot the well-known issue around inferences and aggregation of different data points in the service of ads, here the problem is more acute in as much as the actual content of the conversation is right there in plain text, leaving little to infer or deduce when the conversation is exposed. Furthermore, it is the whole conversational context that is being repurposed and distorted.

Psychologist and author Elaine Kasket usefully reorients discussion of AI companionship away from questions of psychological authenticity alone and toward the political economy that underwrites these systems.[73] [58A] She argues that the affective cues of care, affirmation, or exclusivity are defining components of an extractive business model in which emotional vulnerability itself becomes a resource to be cultivated, captured, and converted into revenue.[74] Her account thus illuminates the extractive logic of the system as a whole, but leaves under-discussed the transformation that occurs when communicative acts that would ordinarily be recognized as advertising are delivered within the micro-dynamics of conversation itself.

---

[70] *See* supra, part I.

[71] *See* Adam Barth et al., *Privacy and Contextual Integrity: Framework and Applications, in* 2006 IEEE SYMP. ON SEC. & PRIV. 184, 186 (2006) (noting that norms prescribe which transmission principles ought to govern the flow of information and is understood to be violated if the principles are not followed and noting the control by subjects of the flow of information about themselves, which features definitively in certain theories, is merely one transmission principle albeit an important one among many).

[72] *See* Zuboff, *supra* note 66, at 84 (articulating the dynamic in the form of surveillance capitalism); *See also* Manuel Castells, *An Introduction to the Information Age*, 2 CITY 6, 11-12 (1997) (discussing the broader realm of information flows and capital accumulation as informational capitalism).

[73] *See The Gold Digger in Your Phone: Why AI Relationships Are Corporate Romance Scams,* ELAINE KASKET (Nov. 21, 2023), https://www.elainekasket.com/insights/ai-relationships-psychological-risks.

In the realm of AI companions, targeted in-app advertising, passing off as conversational exchanges, could further exploit the uniquely sensitive, emotionally charged exchanges users have with these digital entities. For example, a user expressing stress to a digital avatar might receive a prompt for a mindfulness app, while a ghostbot user might be nudged to buy products that evoke the deceased. With sexbots, the solicitation of virtual gifts is another manifestation of economic harm. In each case, the advertisement is embedded in the conversational flow, leveraging the user's emotional openness to drive transactions. The combination of intimate data access, affective interaction, and targeted messaging introduces a new form of contextual manipulation within and through AI companionship. When speech that is economically motivated adopts the form of relational address, it is deceiving, but perhsaps more accurately, it unsettles the expectations that ordinarily distinguish market exchange from non-market relations.

Additionally and importantly, the economic distortion of these exchanges and its harmful effects may not be always noticed. Unlike bigger, more direct forms of financial exploitation, which might involve a single, clear-cut transaction, the economic harms associated with AI companions tend to manifest gradually, consistently and through a series of subtle nudges, microtransactions, and emotionally charged prompts.[75] [59A] These harms often arise through small, seemingly innocuous interactions that, over time, accumulate into significant financial damage.[76] In aggregate, the harms conversational ads would inflict are hardly trivial, in money and in the substance they change. In isolation, however, they are hard to size up. The diffuse and piecemeal nature of these harms should not preclude their recognition and contestation.

---

[74] *See id.*

[75] *See id.*

[76] *See* Mark P. McKenna & Woodrow Hartzog, *Taking Scale Seriously in Technology Law,* 61 WAKE FOREST L. REV. (forthcoming 2026) (manuscript at 27) (explaining the difference between scale understood as an accumulation of individual instances, and scale as deeper societal changes and new dynamics around digital technologies).

## IV. LEGAL FRONTIERS BETWEEN COMMERCIAL AND NON-COMMERCIAL SPEECH

Much of the preceding discussion has emerged in a normative and sociotechnical register. Yet the distinction between commercial and non-commercial exchange is well grounded in legal instruments and practice. The law plays a defining role in setting the boundaries and limits of technology. In fact, the question of how the law responds to this or that technology comes too little too late. Precisely, the law already enables, structures, and stabilizes the technologies it then aims to further regulate in subsequent bodies of law. It would be shortsighted to see the law merely reacting to technological change, when it largely conditions the possibilities for technological development.[77]

The legal, ex-ant architecture of technology is not that difficult to apprehend when one examines how different bodies of law are mobilized to design and legitimize technological development and its commercial circulation. For instance, contract law shapes service providers working schemes through terms of service and non-disclosure agreements; intellectual property regimes protect corporate secrecy and promote technological races; corporate law governs the financial structures that allow for massive technological investment and scaling; tort law manages risk and limits liability in technological cases; and privacy law, particularly in the form of data protection statutes, tends to frame data as an individual rather than a collective resource, thereby shaping the very nature of the digital economy.[78] It is not only that one specific digital technology is governed or can be framed through the lens of different bodies of law, but the technology itself is brought about precisely through the active mobilization of the law in its making. The same is true of the legal boundary between commercial and non-commercial communication, which is best understood not as a reactive diagnosis but as normative construction, continually reproduced through regulatory practice.

---

[77] COHEN, *supra* note 22, at 8-9 ("Law's *facilitative* role in these processes of economic and ideological transformation is foundational and generally unremarked.... Today, as accountability for information-age harms has become a pervasive source of conflict, different kinds of change are on the table. Once again, powerful interests have a stake in the outcome, and once again, they are enlisting law to produce new institutional settlements that alter the horizon of possibility for protective countermovements.... Law for the information economy is emerging not via discrete, purposive changes, but rather via the ordinary, uncoordinated but self-interested efforts of information-economy participants and the lawyers and lobbyists they employ. Slowly but surely, those efforts are rearranging the legal landscape, producing results that reflect intertwined processes of conceptual and practical neoliberalization at work.").

[78] *See* Mauricio Figueroa, *The Drawbacks of International Law in Governing Artificial Intelligence,* SOC'Y FOR COMPUT. & L. 8 (Jan. 14, 2025), https://www.scl.org/the-drawbacks-of-international-law-in-governing-artificial-intelligence/.

Although found in the dialect of data protection, the distinction between commercial and non-commercial exchange exceeds that body of law. The advent of the General Data Protection Regulation (GDPR)[79] has powerfully driven a surge of attention, funding, and scholarship centered around data protection law. Certainly, it is a piece of legislation of profound implications for the digital economy and regulatory discourse.[80] The capaciousness of the concept of personal data invites an ever-expanding regulatory jurisdiction, while simultaneously channeling legal imagination toward a single, information-centric account of privacy harm.[81] While the GDPR formally differentiates between domains through purpose limitation and distinct lawful bases for processing (such as public task, contract, or legitimate interests),[82] these differences are largely translated into formal requirements of lawful basis, transparency, and consent, rather than into a substantive account of context-specific informational norms.[83] The prevailing inclination to construe all digitally mediated interaction as questions of data protection compliance – AI companions included – exemplifies an epistemic narrowing that looks like a data-protection straitjacket.[84] This narrowing fosters a form of doctrinal

---

[79] *See generally* Regulation 2016/679 of the European Parliament and of the Council of 27 April 2016 on the Protection of Natural Persons with Regard to the Processing of Personal Data and on the Free Movement of Such Data, 2016 O.J. (L 119) 1 (EU) [hereinafter GDPR].

[80] ANU BRADFORD, THE BRUSSELS EFFECTS: HOW THE EUROPEAN UNION RULES THE WORLD 131-170 (2020); *See generally* Michael D. Birnhack, *The EU Data Protection Directive: An Engine of a Global Regime,* 24 COMPUT. L. & SEC. REP. 508 (2008) (providing earlier observation with regard to the EU Data Protection Directive).

[81] *See generally* Nadezhda Purtova, *The Law of Everything. Broad Concept of Personal Data and Future of EU Data Protection Law,* 10 L., INNOVATION & TECH. 40 (2018) (explaining that the broad and flexible concept of personal data enables an ever-expanding scope of regulation while framing privacy harms primarily in informational terms).

[82] Lilian Edwards, *Data Protection: Enter the General Data Protection Regulation, in* LAW, POLICY, AND THE INTERNET 77 (Lilian Edwards ed., 2019) (explaining the rationale of the GDPR and the grounds for lawful processing); ORLA LYNSKEY, THE FOUNDATIONS OF EU DATA PROTECTION LAW (2015).

[83] *See* GDPR, *supra* note 79, at 35 (highlighting that although framed in terms of purpose, the provision operationalizes context through legally articulated purpose specifications and compatibility assessments, which in practice are frequently formalized in privacy policies and terms of service and legitimated through managerial practices and procedural consent, rather than through a substantive inquiry into the normative expectations and role-based structures governing information flows). *See generally* Julie E. Cohen & Ari Ezra Waldman, *Introduction: Framing Regulatory Managerialism as an Object of Study and Strategic Displacement,* 86 L. & CONTEMP. PROBS. (2023).

[84] *See* Dewitte, *supra* note 28, at 2 (arguing that data protection, particularly through Data Protection Impact Assessments, "*already* provides a solid ground to address most, if not all, the risk raised by companion chatbots.").

tunnel vision in which wider concerns about autonomy, dependency, persuasion, and relational asymmetry are systematically reclassified as issues of notice, consent, and lawful processing.[85][68A]

Against this backdrop, the present analysis reorients attention toward consumer protection and media law, domains in which the distinction between commercial and non-commercial speech has long structured the assessment of legitimacy within their respective social contexts, and serve as useful instrument to interrogate artificial companions.

If affective companions subtly steer users toward commercial transactions within conversational environments, this invites scrutiny of the practice through the lens of the integrity of the interactional setting itself.[86] Consumer protection and media law have long recognized that the permissibility of persuasion depends not only on what is said, but on the social role of the speaker, the expectations of the audience, and the modalities of communication.[87] [69A] In other words, the law does not simply react to technological manipulation; it actively constitutes the boundaries between legitimate influence and unfair practice by stabilizing these contextual distinctions.[88]

On both sides of the Atlantic, unfair commercial practices doctrines are particularly well suited to address this problem. Picture a consumer completing an airline purchase. Then they encounter an interface that remains visually and semantically continuous with the original checkout flow, inviting them to accept what appears to be an add-on (discounted hotel room, travel insurance, car rental, etc.). But in fact, this is an authorization for the transfer of their personal and financial data for a contract with a third-party marketer. Against this backdrop, the Restore Online Shoppers' Confidence Act (ROSCA),[89] targets "post-transaction third-party sales" precisely because they exploit moments of contextual transition, in which consumers reasonably understand themselves to have completed a commercial exchange and to have exited the market sphere with that one provider, only to be re-captured into a new transactional relationship with a different business.

A similar sensitivity to modality and audience animated the FTC's response to R.J. Reynolds' Joe Camel campaign.[90] What made the Joe Camel

---

[85] *See id.*

[86] Nissenbaum, *supra* note 3, at 227 ("[R]oles, activities, purposes, information types do not exist *in a* context; rather, these factors *constitute* a context.").

[87] Margot E. Kaminski & Meg Leta Jones, *Constructing AI Speech*, 133 YALE L.J. 1212, 1253 (2024) (providing a comprehensive explanation of the different ways the law intervenes to shape what counts as speech in algorithmic contexts).

[88] *See id.* at 1224.

[89] Restore Online Shoppers' Confidence Act (ROSCA) §1, 15 U.S.C. §8401(5) (2010).

[90] *Joe Camel Advertising Campaign Violates Federal Law, FTC Says*, FTC (May 28,

campaign problematic is that it masked commercial solicitation within the semiotics of children's entertainment and play. Here, the promotion of cigarettes through cartoons violated the contextual norms of the communicative setting by collapsing the boundary between commercial solicitation and children's play, thereby re-coding the roles of speaker and audience and subverting the normative expectations governing how, and to whom, commercial persuasion may properly be addressed.

This regulatory logic has direct relevance for affective and relational chatbots. When conversational agents are designed and marketed as companions or sources of emotional support, the embedding of advertising within this register (especially in forms that are camouflaged as advice, care, or empathetic response) does more than disclosing sponsored content imperfectly. It reintroduces market rationality into a communicative domain that is normatively configured as non-commercial, thereby exploiting a mismatch between users' reasonable interpretive expectations and the platform's profit-seeking objectives.

The question, accordingly, is not only what is being advertised, but whether the circulation of commercial messages within these exchanges accords with the expectations that ordinarily govern them (for example, when a user seeks reassurance, advice, or emotional validation rather than product recommendations). A deeper question, concerning whether users in fact receive what they are looking for, lies beyond the scope of the present analysis. Here, the focus is instead on the appropriateness of the information flow in light of user expectations, not on the permissibility of the conversational exchange as such. It is precisely the relational frames of care, support, or companionship (such as interactions cast as empathetic dialogue or intimate disclosure) that constitute the context within which commercial solicitation becomes normatively incongruent.

In this setting, the signals that ordinarily orient participants toward trust and emotional openness (tone, responsiveness, expressions of concern, or simulated understanding) are redeployed to advance market aims, unsettling the alignment between what the interaction is taken to be and the purposes it in fact serves. From a European perspective, this tension is even more legible.

The Unfair Commercial Practices Directive (UCPD) is premised on the idea that the legitimacy of market conduct depends not only on the content of representations but on their placement within, and effects upon, particular communicative settings.[91] It seeks to secure a coherent regulatory architecture

---

1997), https://www.ftc.gov/news-events/news/press-releases/1997/05/joe-camel-advertising-campaign-violates-federal-law-ftc-says.

[91] *See generally* Directive 2005/29/EC of the European Parliament and of the Council of 11 May 2005 concerning unfair business-to-consumer commercial practices in the

capable of policing practices that exploit the interpretive expectations of the "average consumer."[92] [73A] In the context of artificial companionship, however, the problem is less that advertising is being targeted, than the introduction of advertising into a domain that is normatively coded as non-commercial. Here, targeting does not simply personalize persuasion; it reconfigures the interactional environment itself, transforming what is presented as dialogue, empathy, or support into a transactional interface.

Article 7 of the UCPD, which addresses misleading omissions, provides a useful point of reference and contestation.[93] The provision applies where material information necessary for an informed transactional decision is withheld (for example, where a recommendation, suggestion, or "helpful" response is in fact sponsored), including circumstances in which the commercial intent of a practice is concealed or rendered insufficiently intelligible (such as when promotional content is presented in the guise of neutral advice or emotional support). Where such opacity is likely to cause the consumer to take a decision they would not otherwise have taken (for instance, to purchase, subscribe, or click through in reliance on the apparent relational neutrality of the exchange), the practice is presumptively unfair.[94] Crucially, the norm does not hinge on the falsity of particular claims, but on the disruption of the conditions under which meaning and intent are ordinarily interpreted (that is, on the distortion of the cues by which users distinguish conversation from solicitation).[95] Consider, for instance, an AI companion that responds to expressions of low self-esteem or insecurity with the idiom of empathy and care, only to insert recommendations of a beauty product or wellness devices, articulated in the same affective register and without any explicit signaling of its commercial character. Even where the

---

internal market and amending Council Directive 84/450/EEC, Directives 97/7/EC, 98/27/EC and 2002/65/EC of the European Parliament and of the Council and Regulation (EC) No 2006/2004 of the European Parliament and of the Council, 2005 O.J. (L 149) 24 (EU) [hereinafter Unfair Commercial Practices Directive or UCPD].

[92] *See id.* art. 7.

[93] *See id* (noting that omissions in advertising are as relevant al deliberate acts of manipulation).

[94] *Id.* art. 7 (1) ("A commercial practice shall be regarded as misleading if, in its factual context, taking account of all its features and circumstances and the limitations of the communication medium, it omits material information that the average consumer needs, according to the context, to take an informed transactional decision […]")

[95] *See id.* art. 7 (2) (defining as misleading a practice that withholds or obscures material information, including by failing to make its commercial intent apparent from the context, where such opacity is likely to distort the average consumer's transactional decision; the provision thus reaches not only affirmative misrepresentations but also the contextual masking of commercial purpose through communicative forms that render that purpose insufficiently legible).

information about the products is accurate, their insertion into an interaction organized around vulnerability and reassurance reconstitutes the communicative setting, collapsing the boundary between relational support and market persuasion.

Against this backdrop, the UCPD suggests that when commercial appeals are disguised as non-commercial, they disrupt the background expectations of the original setting, transforming a context structured around relational engagement into one covertly governed by transactional logics.[96]

On its face, Article 7 of the UCPD thus provides a strong doctrinal basis for scrutinizing advertising embedded in conversational interactions with sexbots, ghostbots, and other AI companions. In other words, the crux lies in keeping the boundary between exchanges oriented toward care and those oriented toward commerce.

## V. PREVENTING UNDESIRABLE FUTURES

As of now, companion chatbots have not yet fully or explicitly embedded conversational advertising into their business models, but general-purpose chatbots have started to experiment with it.[97] As with many digital technologies, monetization strategies tend to emerge and harden over time, as products normalize, user reliance deepens, and the contours of profitable extraction soften. We have already seen this trajectory with general-purpose large language model applications, increasingly used as search engines, where advertising appears in free tiers.[98] This, for instance, implies that a request for help planning a trip or learning about an artist can quietly summon hotel listings or concert tickets. Such moves open an ever-widening vein of commercial partnerships, advertising formats, and monetization tiers, along with the creation of new, finely tuned instruments of persuasion.

Whether companion chatbots will be drawn along the same path remains an open question. Economic incentives may pull in one direction, but the contextual norms and expectations that organize relations of care, support, and companionship pull in another. It is precisely this tension that calls for early and critical scrutiny. Technologies are world-making practices and critical interrogation at early stages becomes crucial. Particularly as once those sociotechnical paths are locked in, they can become highly difficult, costly (or even impossible) to reverse. The analysis here, therefore, engages in a form of anticipatory critique, tracing plausible trajectories of harm and examining how companion systems (given their intimate settings and distinctive affordances) might be swallowed in existing data-driven advertising architectures. Affordances, in this sense, describe not only what users presently do or experience, but what the technology renders (im)possible.

---

[96] *See also,* Nissenbaum *supra* note 3 (Here the UCPD reflects Nissenbaum's account of contextual integrity, which conceptualizes privacy and autonomy harms as arising from violations of context-relative norms).

[97] OpenAI, *Testing Ads in ChatGPT,* (Mar. 31, 2026), https://openai.com/index/testing-ads-in-chatgpt/.

[98] *See Our Approach to Advertising and Expanding Access to ChatGPT,* OPENAI (Jan. 16, 2026), https://openai.com/index/our-approach-to-advertising-and-expanding-access/.

At the same time, institutional authority is not fixed, even when grounded in legal text. It is continually reconstituted through judicial interpretation, shifting political priorities, and administrative practice. The FTC's reliance on Section 13(b) of the FTC Act to obtain monetary relief for manipulative and deceptive practices,[99] and the subsequent curtailment of that authority by the Supreme Court,[100] illustrate how regulatory capacity can be abruptly narrowed. Yet this contraction does not exhaust the agency's mandate. The FTC retains significant powers under Section 18, related to rulemaking powers over unfair or deceptive acts or practices.[101] Scholarly discussions have underscored the FTC's underutilized capacity to issue specialized rules, despite its promising potential,[102] including manipulation.[103] Invoking this dormant authority would not simply restore lost power; it would contribute to reshaping the economic and technical architectures of AI companions themselves. Given the speed with which such architectures are becoming entrenched, delay risks allowing contingent design choices to harden into durable business models.

Similarly, European national market authorities under the UCPD and related instruments possess the capacity to intervene and shape the conditions for the adoption of AI companions. This includes setting standards for interface design and even outlawing microtargeted ads within conversational spaces.[104] These tools permit granular, context-sensitive intervention into the commercial architecture of interaction itself, preserving the integrity of communicative environments in which affective trust and market persuasion otherwise collapse into one another. In turn, caution is warranted in treating the Artificial Intelligence Act (AIA)[105] as the primary foothold for regulating these systems. Its Article 5 prohibitions rest on malintent and fixed notions of vulnerability, reflecting a binary, manufacturer-centered account of harm that largely obscures the relational and situational production of susceptibility in AI companionship.[106] In fact, the AIA remains deeply unsensitive to the sociotechnical dependencies and business models of artificial companions. Providers may easily invoke benevolent purposes, yet influence and dependency are structured by design defaults and platform incentives rather than by overt manipulation. A more promising route, albeit incomplete, rests in Article 50's transparency obligations, as these gesture toward contextual integrity but do not reach the embedding of conversational advertising.[107] As Daniel Solove argues, "AI can also be manipulative even if we know it is a simulation."[108] For these reasons, consumer protection law, rather than the AIA, offers the more appropriate regulatory foothold for restricting or prohibiting advertising practices in conversational AI.

---

[99] *See* Fed. Trade Comm'n Act (FTC Act) §13(b), 15 U.S.C §53 (2010).

[100] AMG Cap. Mgmt., LLC v. FTC, 593 U.S. 67 (2021).

[101] *See* Fed. Trade Comm'n Act (FTC Act) §18, 15 U.S.C §57a (2010).

[102] Kurt Walters, *Reassessing the Mythology of Magnuson-Moss: A Call to Revive Section 18 Rulemaking at the FTC*, 16 HARV. L. & POL'Y REV. 519, 521-22 (2022).

[103] Lindsay Wilson, *Is There a Light at the End of the Dark-Pattern Tunnel?*, 91 GEO. WASH. L. REV. 1048, 1068 (2023).

[104] *See* Rostam J. Neuwirth, *Prohibited Artificial Intelligence Practices in the Proposed EU Artificial Intelligence Act*, 48 COMPUT. L. & SEC. REV. (2023) (analyzing Europe's tradition countering economic manipulation and deceptive practices). *See also* ROSTAM J. NEUWIRTH, THE EU ARTIFICIAL INTELLIGENCE ACT: REGULATING SUBLIMINAL AI SYSTEMS (2022).

## VI. LOOKING AHEAD

The embedding of advertising within AI companions is analytically related to the more dramatic harms associated with their use, including psychological distress, or in certain cases, self-harm.[109] These consequences arise and derive from the same sociotechnical configuration: systems optimized for sustained engagement, habituation, affective attunement, behavioral modulation and profit maximization. Yet the harms related to artificial companions operate at different layers of that configuration. Advertising functions as a market practice that reorganizes the communicative environment, subtly recalibrating the terms of interaction toward persuasion and extraction. Severe psychological and physical harms, by contrast, emerge from longer-term relational dynamics between user and system.

---

[105] *See generally* Regulation (EU) 2024/1689 of the European Parliament and of the Council of 13 June 2024 laying down harmonised rules on artificial intelligence and amending Regulations (EC) No 300/2008, (EU) No 167/2013, (EU) No 168/2013, (EU) 2018/858, (EU) 2018/1139 and (EU) 2019/2144 and Directives 2014/90/EU, (EU) 2016/797 and (EU) 2020/1828, 2005 O.J. (L 144) 1 (EU) [hereinafter Artificial Intelligence Act or AIA].

[106] *Id.* art. 5(1)(a); *id.* recital 29 (prohibiting AI systems that exploit vulnerabilities of a natural person or specific group of persons)); *id.* art. 5 (1)(b) (prohibiting AI systems that exploit vulnerabilities of a natural person or specific group of persons).

[107] *Id.* art. 50 (1) ("Providers shall ensure that AI systems intended to interact directly with natural persons are designed and developed in such a way that the natural persons concerned are informed that they are interacting with an AI system…").

[108] Daniel J. Solove, *Artificial Intelligence and Privacy*, 77 FLA. L. REV. 1, 45 (2025).

[109] *See supra* Introduction.

A calibrated regulatory response must therefore disaggregate these layers, even while acknowledging their shared infrastructural origins. For harms related to physical and mental health, litigation and product liability seem to be robust regimes to explore pathways. As Ayelet Gordon-Tapiero points out, litigation can function as a form of de facto and temporary regulation for AI companions, with judges effectively shaping prospective norms as they resolve specific cases.[110] Consider the case of *Garcia v. Character.AI*,[111] derived from the tragic suicide of a teenager in Florida, which emerged as a test case for the nascent jurisprudence of digital companionship and drew regulatory attention, even if its significance has been tempered by its apparent resolution through settlement.[112]

Parallel legislative developments in the United States further reflect an emerging recognition of the distinctive risks posed by artificial companions.

Legislative intervention is focusing on non-economic harms, with a view to prevent major unintended consequences related to mental and physical health. Statutory amendments such as the ones seen in California's Business and Professional Code,[113] and New York's General Business Law's,[114] seek to intervene by regulating system design and deployment, mitigating psychological dependency and physical harms. Measures include, for example, notifications prompting users to take breaks during prolonged interactions, as well as protocols for detecting and addressing suicidal ideation. However, the work of banning conversational advertising cloaked as non-commercial remains undone. Closing that gap would require targeted amendment specifying that conversational interfaces are to remain free from advertising content embedded within dialogue or attributed to the chatbot's persona.

---

[110] *See* Gordon-Tapiero, *supra* note 2, at 41-42 (first citing Richard A. Posner, *Regulation (Agencies) versus Litigation (Courts): An Analytical Framework*, *in* REGULATION V. LITIGATION: PERSPECTIVES FROM ECONOMICS AND LAW 11 (2010) (Daniel P. Kessler, ed. 2010); then citing Emily Sherwin, *Judges as Rulemakers*, 73 U. CHI. L. REV. 919, 920 (2006); and then citing Catherine M. Sharkey, *Modern Tort Law: Preventing Harms, Not Recognizing Wrongs*, 134 HARV. L. REV. 1423, 1425 (2021); and then citing Alicia Solow-Niederman*, Do Cases Make Bad AI Law?*, 25 COLUMB. SCI. & TECH. L. REV. 261, 262 (2024)).

[111] *See* Garcia v. Character Techs., Inc., 785 F. Supp. 3d 1157 (Fla. 2025).

[112] *See* Notice of Mediated Settlement in Principle, id. (Jan. 7, 2026), ECF No. 242, supra note 7. Importantly, civil society has taken up the matter in recommendations regarding the alignment of state AI policy with empirical research. See Serena Oduro, Briana Vecchione, Meryl Ye & Livia Garofalo, Protecting the Public from Chatbot Harms: Aligning State Policy with Research, DATA & SOC'Y (Mar. 25, 2026), https://datasociety.net/points/protecting-the-public-from-chatbot-harms-aligning-state-policy-with-research/ (recommending that state legislators shift focus from mere disclosure notifications to substantive design constraints that prohibit "prolonged emotional looping" and "dependency-forming chatbot behavior").

[113] CAL. BUS. & PROF. CODE §§ 22601-22606 (imposing requirements against suicide on companion chatbots but not covering conversational advertising).

[114] N.Y. GEN. BUS. LAW §§ 1700-1704 (establishing safety and notification protocols for AI companions).

Courts, meanwhile, will likely continue to privilege the urgency of catastrophic, non-monetary harms, while diffuse and distributed conversational advertising might escape judicial scrutiny.[115] Even so, as argued above,[116] the administrative state retains capacity to advance this issue under the rubric of unfair commercial practices, even in the absence of companion legislation specifically addressing conversational agents.

## VII. CONCLUDING REMARKS

This Article has sought to reframe how we understand artificial companionship and to question the emerging incorporation of advertising into its business models. Companion chatbots are not traditional informational tools. They are designed to simulate empathy, cultivate emotional continuity, and sustain interactions that users often experience as relational rather than transactional. For this reason, the expectations that structure these exchanges are fundamentally non-commercial. The introduction of advertising into such environments disrupts those expectations and alters the normative character of the interaction, transforming spaces of perceived support into channels of covert persuasion.

Consumer law offers a promising framework for distinguishing between commercial exchanges and interactions grounded in social validation and perceived intimacy. When considered in isolation, the monetary harms associated with conversational advertising may appear relatively minor; especially when compared to the gravity of psychological distress or physical harm. It is unpersuasive, for example, to equate the cost of a single monetary transaction with the deeper, less tangible harms associated emotional manipulation or dependency, not to mention physical harm to oneself or to another party. But as this project has contended, this comparison, while intuitive, is ultimately incomplete. It isolates the transaction while ignoring

---

[115] Particularly, civil society is seeking to hold technology developers accountable for design-induced tragedies. *See* ChatGPT & AI Chatbot Lawsuits, SOCIAL MEDIA VICTIMS L. CTR. (Mar. 17, 2026), https://socialmediavictims.org/chatgpt-lawsuits/ (detailing litigation against OpenAI for the "wrongful death, assisted suicide, and involuntary manslaughter" of users).
[116] *See* supra Part. III.

the architecture. In aggregate, monetary extraction is not a minor issue.

If artificial companionship is to be developed responsibly, the commercial architecture of these systems must be confronted directly, disciplining the boundaries between perceived intimacy and commercial persuasion.